\def\year{2015}
\documentclass[letterpaper]{article}
\usepackage{flushend}
\usepackage{aaai}
\usepackage{courier}
\usepackage{booktabs}
\usepackage{url}
\usepackage{array}
\usepackage{amsmath}

\usepackage{latexsym}
\usepackage{color}
\usepackage[usenames]{xcolor}
\usepackage{amssymb}
\usepackage{rotating}
\usepackage{multirow}
\usepackage{pdfsync}
\usepackage{url}
\usepackage{enumitem}

 \newif\ifdraft
\drafttrue

\begin{document}
%
\title{Trump vs. Hillary\\Analyzing Viral Tweets during US Presidential Elections 2016}
\author{Walid Magdy$^1$ and Kareem Darwish$^2$\\
$^1$School of Informatics, The University of Edinburgh, UK\\
$^2$Qatar Computing Research Institute, HBKU, Doha, Qatar\\
Email: wmagdy@inf.ed.ac.uk, kdarwish@qf.org.qa\\
Twitter: @walid\_magdy, kareem2darwish\\
}
\maketitle


\begin{abstract}
\begin{quote}
In this paper, we provide a quantitative and qualitative analyses of the viral tweets related to the US presidential election. 
In our study, we focus on analyzing the most retweeted 50 tweets for everyday during September and October 2016. The resulting set is composed 3,050 viral tweets, and they were retweeted over 20.5 million times.  We manually annotated the tweets as favorable of Trump, Clinton, or  neither. Our quantitative study shows that tweets favoring Trump were usually retweeted more than pro-Clinton tweets, with the exception of a few days in September and two days in October, especially the day following the first presidential debate and following the release of the Access Hollywood tape. On two days in October 2016, pro-Trump tweet volume 
accounted for than 90\% of the total tweet volume. 
%
%
%
%
\end{quote}
\end{abstract}



\section{Introduction}
\label{sec:intro}
Social media is an important platform for political discourse and political campaigns \cite{shirky2011political,west2013air}. Political candidates have been increasingly using social media platforms to promote themselves and their policies and to attack their opponents and their policies.  Consequently, some political campaigns have their own social media advisers and strategists, whose success can be pivotal to the success of the campaign as a whole. In the context of this paper, we are interested in measuring the volume and diversity of support for the two main candidates for the 2016 US presidential elections, Donald Trump and Hillary Clinton, on Twitter during the two months preceding the elections, namely September and October 2016. The work is based on the data being collected via \url{TweetElect.com}, which is an online website that tracks tweets and Twitter trends pertaining to the US presidential elections.

For our analysis, we use the top 50 retweeted tweets, aka viral tweets, for everyday in September and October 2016 pertaining to the US presidential election. The total number of unique tweets that we analyze is 3,050, whose retweet volume of 20.5 million retweets accounts for more than 50\% of the total retweet volume during these two months on TweetElect.  After manually tagging all the tweets in our collection for support for either candidate, we looked at: which candidate has more traction on Twitter and a more diverse support base; when a shift in the volume of supporting tweets happens; and which tweets were the most viral.

We observed that retweet volume of pro-Trump tweets dominated the retweet volume of pro-Clinton tweets on most days during September 2016, and almost all the days during October. A notable exception was the day after the first presidential debate and the day after the leak of the Access Hollywood tape in which Trump used lewd language. 





\section{Data Collection}
In this section, we describe the collection of viral tweets. We initially give an introduction to TweetElect website, which is the source we used to identify the daily viral tweets. Later, we explain the data annotation process and give some statistics on the data collected.

\subsection{TweetElect}
\label{sec:tweetelect}
\url{TweetElect.com}\footnote{\url{http://www.tweetelect.com/}} is a free website that aggregates and shows the most retweeted tweets related to the 2016 US presidential election. The website shows tweets about the elections in general, with the option of displaying tweets about each of the two main candidates separately. It offers search functionality with filters on media type (text, image, video, or links), while enabling the display of search results related to each candidate separately. TweetElect shows the most retweeted tweets, images, videos, and links in the last hour, 12 hours, 1 day, or 2 days.

During September and October, the number of aggregated tweets per day (including retweets) related to the US presidential elections typically ranged between 300k and 600k. This number increased dramatically after specific events or revelations, such as after the presidential debates, where the number of tweets exceeded 4 million tweets.

TweetElect is a special edition of TweetMogaz\footnote{\url{http://www.tweetmogaz.com/}}~\cite{magdy2013tweetmogaz}, which is an Arabic news portal that automatically generates news from tweets. It uses state-of-the-art adaptive filtering methods for detecting relevant tweets on broad and dynamic topics, such as politics and elections~\cite{magdy2016unsupervised}.
TweetElect used an initial set of 38 keywords related to the US elections for streaming relevant tweets. Consequently, adaptive filtering continuously enriches the set of keywords with additional terms that emerge by time~\cite{magdy2016unsupervised}.


\subsection{Tweet Collection}
We were interested in analyzing the most ``viral'' tweets pertain to the US presidential elections.  Therefore, we constructed a set of the most retweeted 50 topically relevant (as provided by \url{TweetElect.com}) tweets for everyday in September and October 2016. Thus, our collection contained  3,050 unique tweets that were retweeted 20.5 million times.  By month, September had 1,500 unique tweets with a retweet count of 6.67 million, and October had 1,550 tweets with a retweet count of 13.89 million. This volume of retweets represents more than 50\% of the total tweet volume collected by TweetElect related to the US elections during September and October 2016.

In our analysis, we show statistics based on three types of viral tweets:
\begin{itemize}
\item \textbf{Top50}: The top 50 viral tweets per day.
\item \textbf{Top10}: The top 10 viral tweets per day. Checking only the top 10 rather than 50 can give a better indicator of the direction of the trends on Twitter on that day, and the top few retweeted tweets usually dominate the retweet volume.
\item \textbf{Top10F}: The top 10 viral tweets per day for the candidates' supporters (``Fans'') only and excluding tweets from the official accounts of the candidates. Since many of the top tweets are usually coming from the presidential candidates, this gives a depth on the support of the candidates by what their fans say. 
\end{itemize}

\begin{figure}[t]
\centering
\includegraphics[width=\columnwidth]{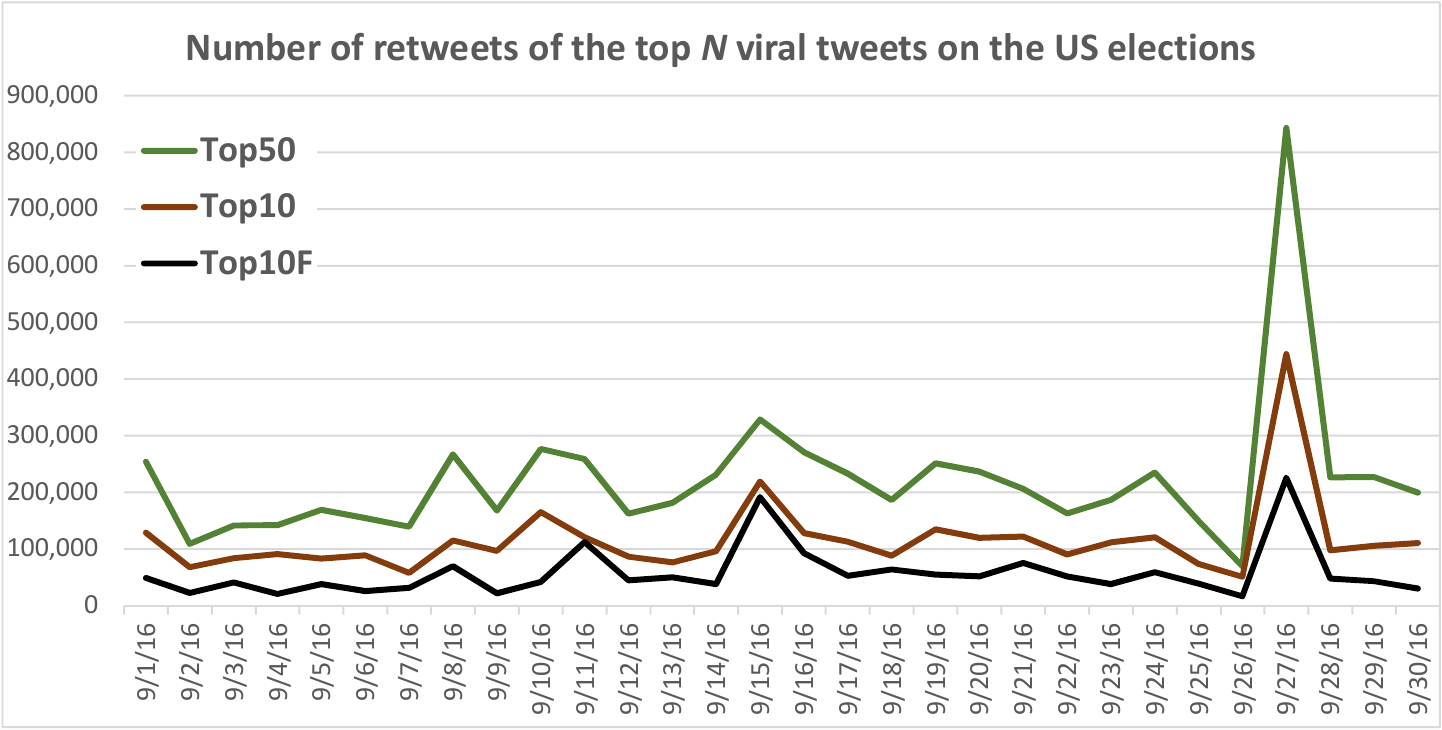}
\caption{\label{fig:RetweetsVolume}Total number of retweets of the Top50, Top10, and Top10F daily viral tweets relating to the US elections during September 2016}
.\\
.\\
\includegraphics[width=\columnwidth]{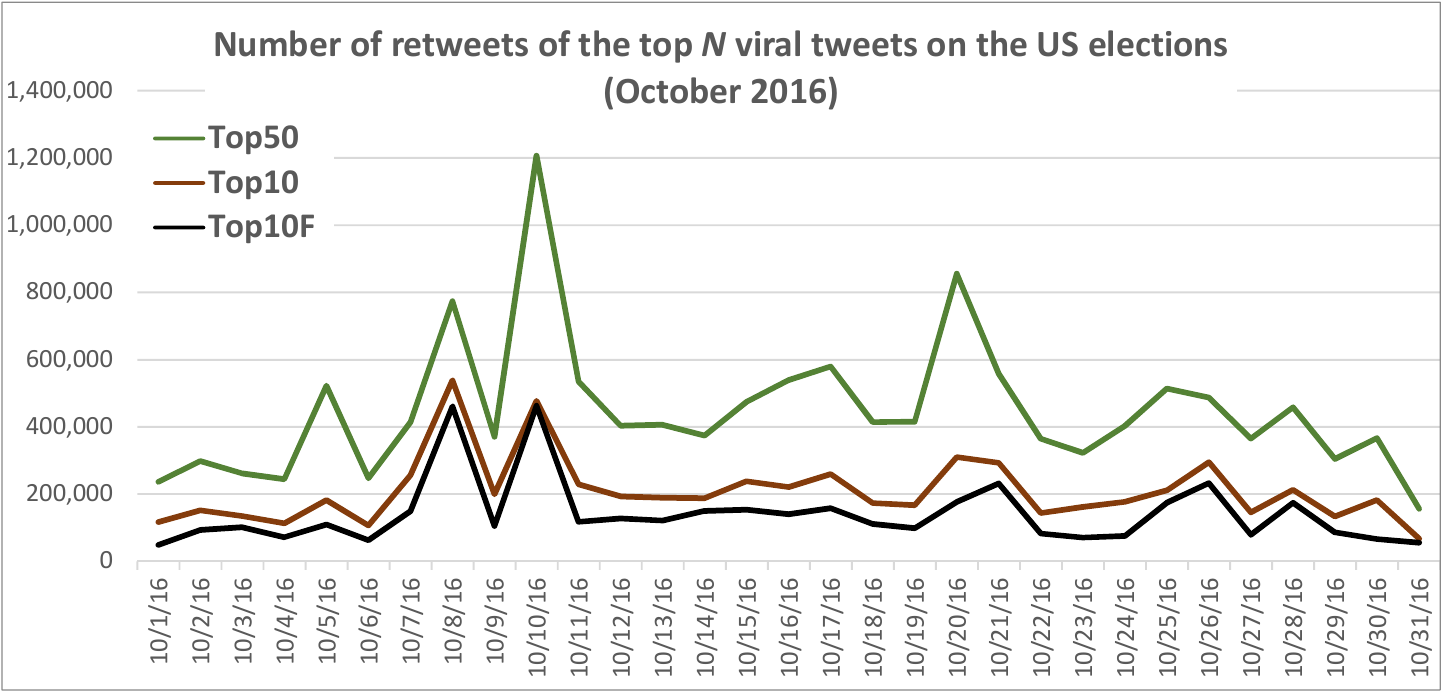}
\caption{\label{fig:RetweetsVolumeOct}Total number of retweets of the Top50, Top10, and Top10F daily viral tweets relating to the US elections during October 2016}
\end{figure}

The total number of retweets for the TOP 50, TOP 10, and TOP 10 Fa	n tweets per day are 6.67 million, 3.49 million, and 1.75 million retweets respectively. Figures \ref{fig:RetweetsVolume} and  \ref{fig:RetweetsVolumeOct} show the virality of the top tweets day-by-day during September and October 2016 respectively. As shown, the days with the largest number of retweets for the top $N$ tweets were September 27 and October 10, which are the days following the first and second debate between the candidates~\footnote{Tweets time stamp is based on GMT}.

\subsection{Tweet Labeling}
We labeled the tweets on two stages, as we labeled the viral tweets of each month directly after the end of this month. Tweets were labeled as pro-Trump, pro-Clinton, or neither.

Out of the 1,500 tweets collected during September, 636 were tweeted by either of the candidates' official accounts. This number was 612 out of 1,550 for October. These tweets were automatically annotated to be in the favor of the candidate who posted them. The remaining tweets were then posted to a crowd-sourcing platform\footnote{\url{https://www.crowdflower.com/}} to be manually annotated.  Each tweet was annotated by at least 3 annotators, and the majority voting is taken for selecting the final label. A golden control set of 17 tweets was provided to control the annotators work quality. 

We asked annotators to label each of the tweet with one of three labels: 1) In favor of Trump, 2) In favor of Clinton, 3) Neither of them.

Instruction were given to annotators as follows:
\begin{itemize}
\item Tweets in favor of a candidate can be:
\begin{enumerate}
\item Clearly showing support to the candidate
\item Giving positive facts about the candidate or his/her campaign (for example showing that he/she leads in polls)
\item Attacking the other candidate
\item Making fun (or sarcasm) of the other candidate or his/her supporters.
\end{enumerate}
\item ``Neither of them'' tweets is selected if:
\begin{enumerate}
\item The tweet is reporting news about the elections or a presidential candidate with no clear bias to be in favor or against.
\item The tweet is attacking both candidates.
\end{enumerate}
\end{itemize}

Annotators were instructed to check the tweet content carefully including any images, videos, or external links to have an accurate annotation. In addition, we advised them to check the profile of tweet authors to better understand their position towards the candidates if needed.

The annotators inter-agreement was 90\%, which is considerably high among three annotators for an annotation task with three choices. This gives high confidence in the quality of annotation.

\section{Results and Analysis}

\begin{figure*}
\centering
\includegraphics[width=0.7\textwidth]{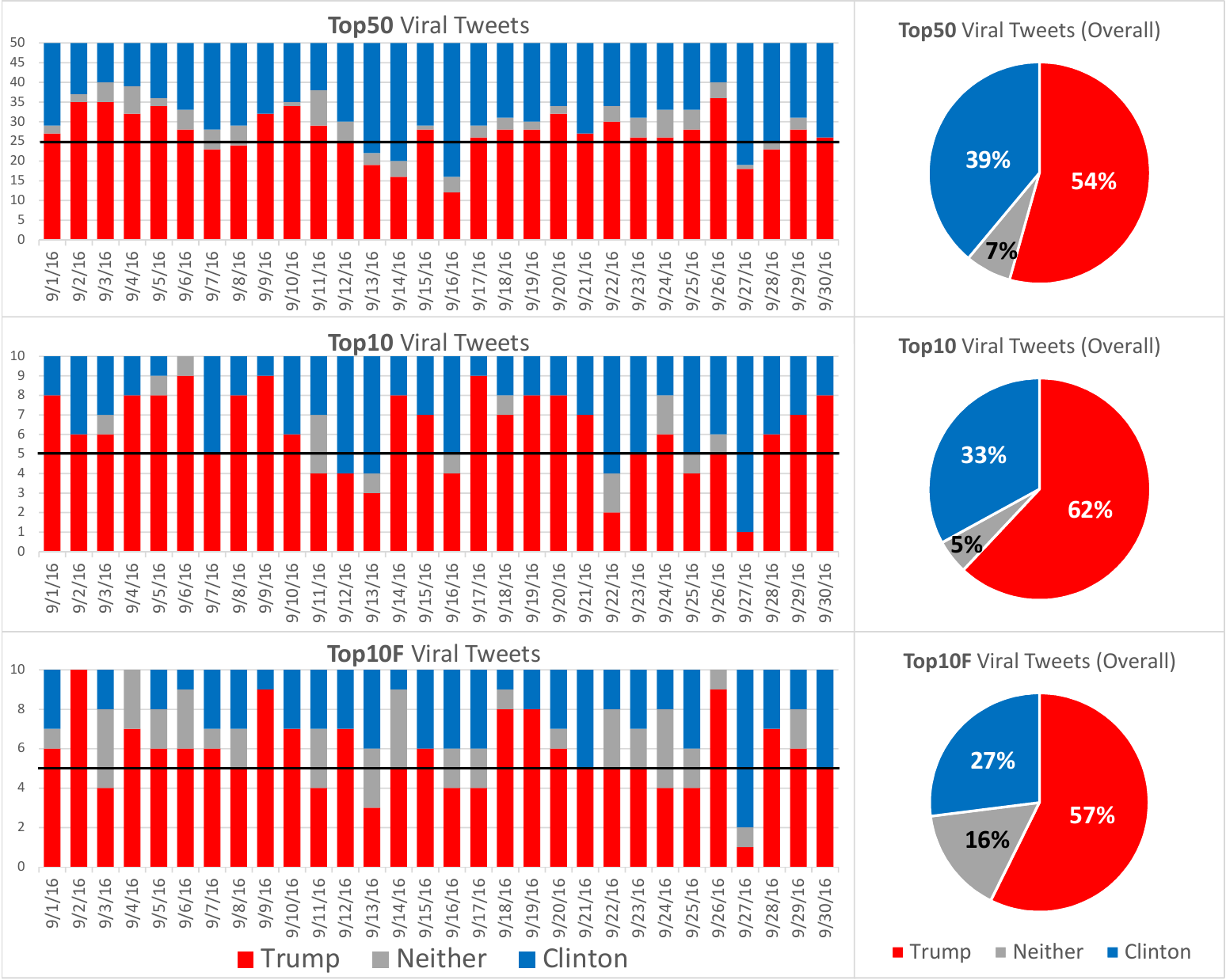}
\caption{\label{fig:ViralTweets}The direction of support of the top viral tweets during September on the US presidential election 2016}
\end{figure*}
\begin{figure*}
\centering
\includegraphics[width=0.7\textwidth]{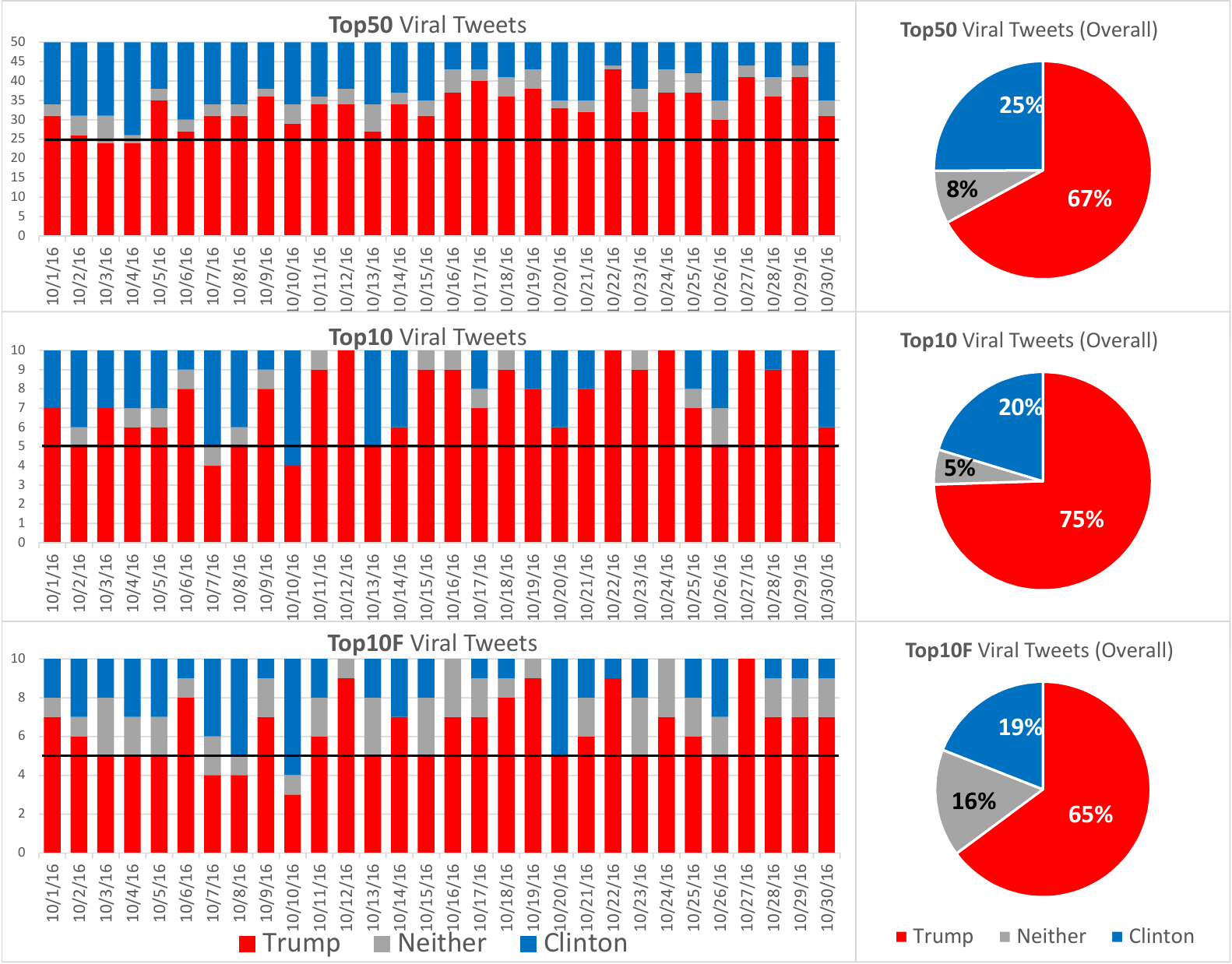}
\caption{\label{fig:ViralTweetsOct}The direction of support of the top viral tweets during October on the US presidential election 2016}
\end{figure*}

Figures \ref{fig:ViralTweets} and \ref{fig:ViralTweetsOct} show the direction of support of the Top50, Top10, and Top10F viral tweets per day for September and October respectively. As it is shown, the number of daily viral tweets in favor of Trump is usually larger than that for Clinton. This observation does not change when considering the Top 50 or 10 tweets. Furthermore, when considering the top viral tweets by their fans (excluding the official accounts of the candidates), the gap between Trump and Clinton was even larger, where 57\% and 65\% of the Top10F tweets in September and October respectively were in favor of Trump, whereas it was only 27\% and 19\% in favor of Clinton respectively. Looking at the daily viral tweets, it is evident that for most days, Trump had more viral tweets supporting him than Clinton. 

For September 2016, Trump led Clinton in the volume of viral tweets everyday during the month save 6 days, with the day following the debate (9/27) showing the highest percentage and volume of tweets supporting Clinton.  We inspected the other 6 days and they corresponded to: (September 12-13) Trump complaining about how the debates would be moderated and questions about Clinton's health; (September 15-16) Trump saying that women are bad for business~\cite{Indepedent} and disavowing the ``birther'' claims that President Obama was not born in the US; and (September 23-24) nothing noteworthy. Trump also led Clinton in the number of unique viral tweets everyday except for 5 days, namely the 2 days following the debate, and on September 12, 13, and 16. 

Also in September, Pro-Trump tweets surged to more than 70\% of the volume of retweets on 3 days, namely September 4, 6, and 9. On September 4, news broke that Clinton mishandled classified material and Trump's top tweet receiving 19.5k retweets which stated: ``Lyin Hillary Clinton told the FBI that she did not know the C markings on documents stood for CLASSIFIED. How can this be happening?'' .  On the 6$^{th}$, hacked Clinton emails were made public with with Trump's top tweet on the topic receiving more than 14k retweets. On the 9$^{th}$, Clinton stated that half of Trump supporters are a ``basket of deplorables''.

For the month of October, pro-Clinton tweet volume exceeded pro-Trump tweet volume on only two days namely Oct. 7 and Oct. 26.  October 7 coincides with the release of the Access Hollywood video of Trump making lewd comments, and coincides with the day when pro-Clinton tweets received the largest portion of retweet volume (58\%).  The top retweeted tweets favoring Clinton were not as much pro-Clinton as they were anti-Trump.  The most retweeted tweet belonged to Jeb Bush, the former hopeful in the Republican presidential primaries, in which he condemned Trump's comments.  Though the tapes were described by some as an ``October surprise'', pro-Trump tweets constituted 35\% of the tweet volume.  On October 26, the user ``Bailey Disler'' was praising the person who destroyed Trump's star on the Walk of Fame.  His tweet was retweeted more than 124k times, which accounts for 6\% of the volume for that day.

Also in October, Pro-Trump retweet volume outpaced pro-Clinton volume for all other days.  In fact, pro-Trump retweet volume accounted for more than 75\% of the volume on 13 day during the month.  Aside from Oct. 29 when the Federal Bureau of Investigation (FBI) announced that they are investigating Hillary Clinton over misuse of email and Oct. 11 when he attacked the leaders of the Republican Party, nothing out of the ordinary was happening any of the other days.  The two most prominent sources of pro-Trump viral tweets were the official accounts of Trump and WikiLeaks.  

\begin{figure}[t]
\centering
\includegraphics[width=\columnwidth]{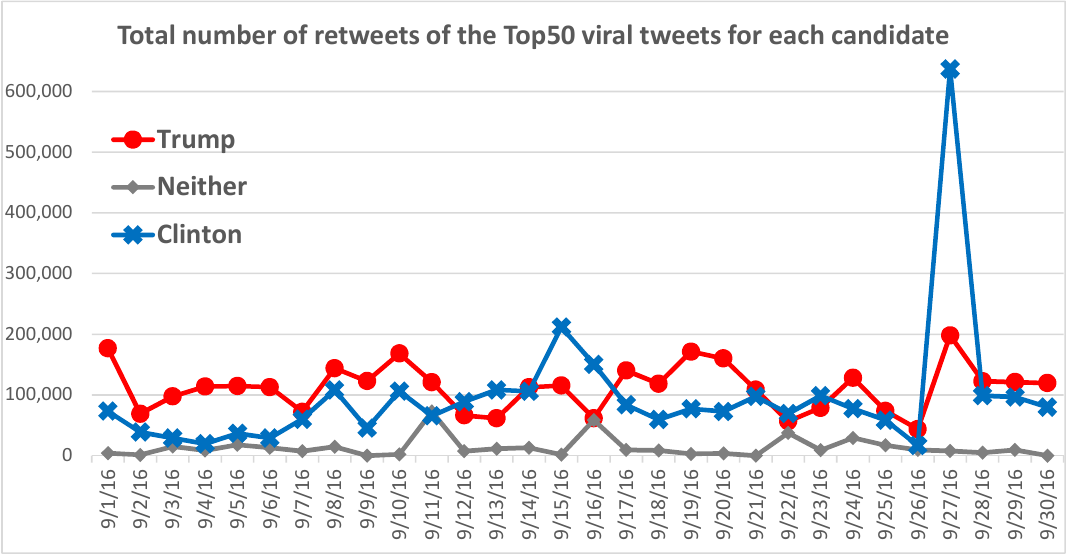}
\caption{\label{fig:ViralTweetsDetails}Total number of retweets of the Top50 daily viral tweets for each candidate during September 2016}
.\\.
\includegraphics[width=\columnwidth]{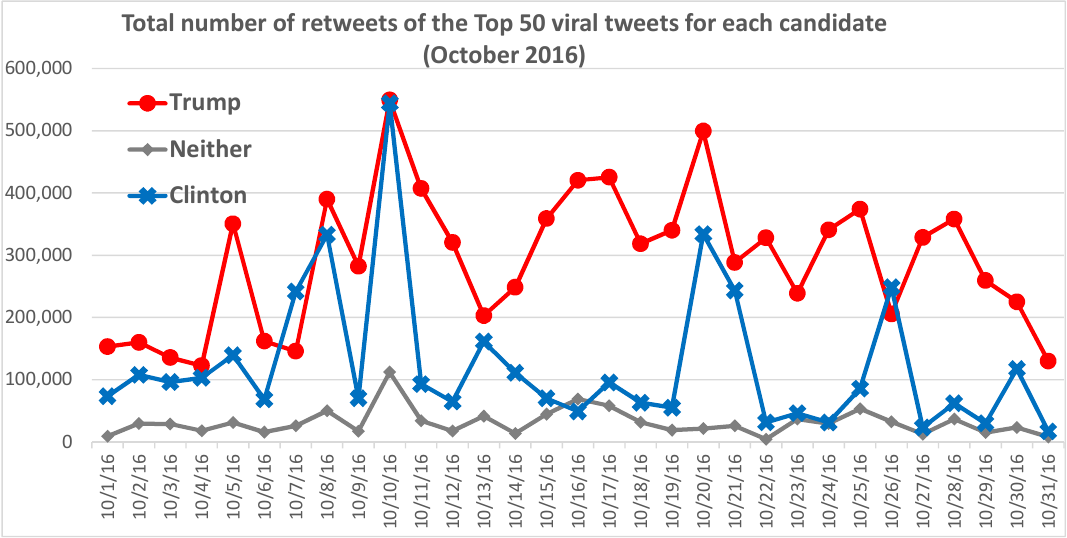}
\caption{\label{fig:ViralTweetsDetailsOct}Total number of retweets of the Top50 daily viral tweets for each candidate during October 2016}
\end{figure}

Results in Figures~\ref{fig:ViralTweets} and \ref{fig:ViralTweetsOct} are based on the top $N$ unique tweets. When we take into consideration the real volume these tweets represents, i.e. counting each tweet in the top 50 with its number of retweets, the results can be seen in Figures~\ref{fig:ViralTweetsDetails} and \ref{fig:ViralTweetsDetailsOct} for the Top50 for September and October respectively. The graphs still show that Trump had a larger number of supporting viral tweets for most of the days. However, when Clinton gets more viral tweets, some of the tweets may receive an enormous number of retweets, as is clear on September 15 and 27 when there were large spikes in the favor of Clinton. The overall share of retweets volume during the month continues to be in favor of Trump. Even with the spikes in pro-Clinton retweets, 54\% and 67\% of the retweet volume for September and October respectively was for tweets supporting Trump, 39\% and 25\% were pro-Clinton, and the remaining 7\% and 8\% were unbiased news or against both candidates.

\begin{figure}[t]
\centering
\includegraphics[width=\columnwidth]{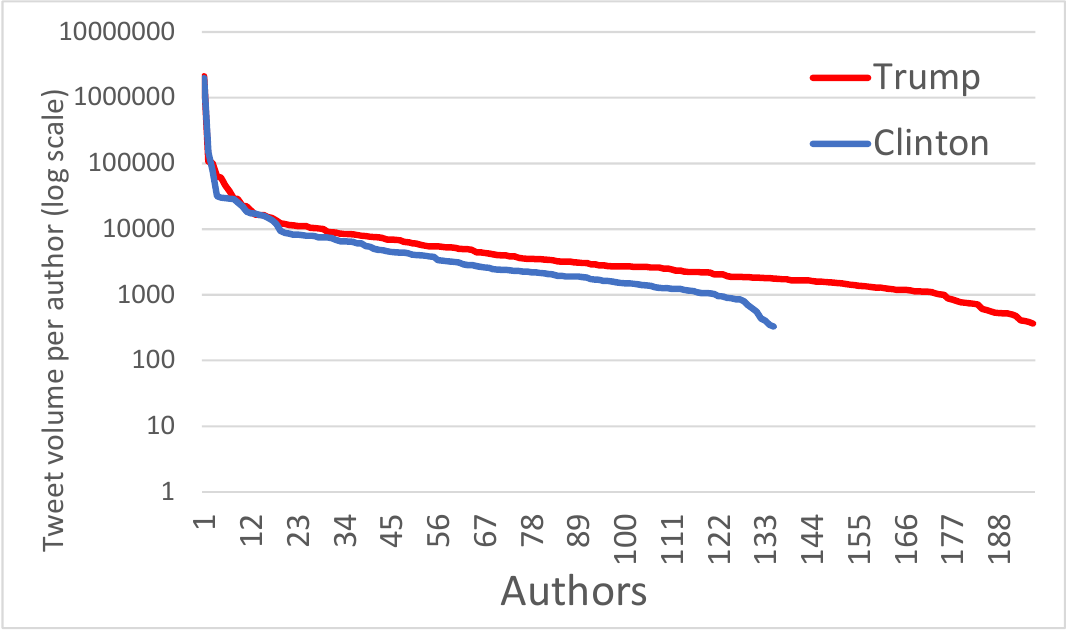}
\caption{\label{fig:VolumeTweetsCounts}Total number of retweets per users whose tweets showed in the Top50 during September 2016}
.\\.
\centering
\includegraphics[width=\columnwidth]{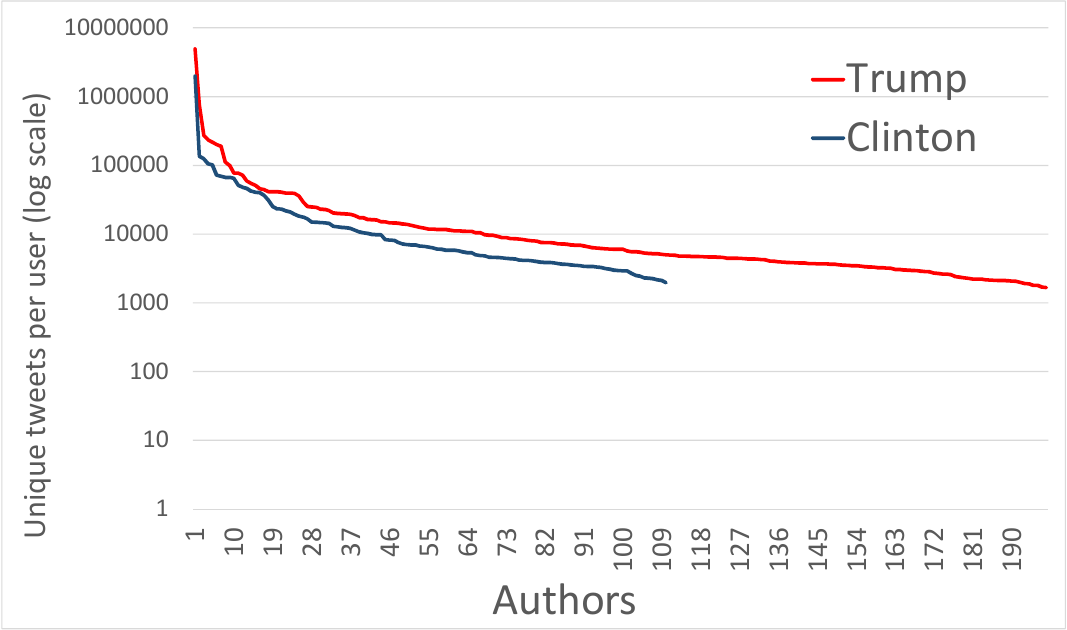}
\caption{\label{fig:VolumeTweetsCountsOct}Total number of retweets per users whose tweets showed in the Top50 during October 2016}
\end{figure}

We also looked at the diversity of tweet authors, and found that there was not only more retweets in support of Trump, but also they were authored by more people.  The viral tweets in support of Trump were authored by 196 and 198 users for September and October respectively compared to 135 and 110 users respectively for Clinton. Furthermore, for September, in terms of the number of unique viral tweets the percentage of viral tweets supporting Trump that were authored by his official account was 31\% compared to 66\% of pro-Clinton tweets that were authored by her official account. The percentages for October were consistent with 35\% for Trump and 63\% for Hillary.  Figures \ref{fig:VolumeTweetsCounts} and \ref{fig:VolumeTweetsCountsOct} show the volume of retweets per author in our set of viral tweets and suggests that there is greater diversity of pro-Trump tweet authors. Conversely, we looked at the 50 tweets with the most retweets over the entire two months of September and October that are supporting either candidate.  For September, 45 were authored by the official Trump account (making up 92\% of the volume of the top 50 tweets) compared to 39 that were authored by the official Clinton account (making up 68\% of the volume of the top 50 tweets). For October, 32 were authored by the official Trump account (making up 57\% of the volume of the top 50 tweets) compared to 23 that were authored by the official Clinton account (making up 39\% of the volume of the top 50 tweets).  The drop in percentages of top retweeted tweets being authored by the candidates from September to October is notable and warrants more investigation. Table \ref{table:TopViralTweeps} and \ref{table:TopViralTweepsOct} show the top 10 most retweeted Twitter accounts along with the number of tweets in our set and retweet volume~\footnote{The number of retweets in Tables~\ref{table:TopViralTweeps}-\ref{table:TopViralTweetsOct} are taken at the time of the study. This number is expected to change over time}.  One interesting observation is that accounts with some of the most retweeted pro-Clinton tweets had only one tweet in our set and there was no correlation between tweet counts and retweet volume.  For accounts with pro-Trump tweets, there is correlation between tweet count and retweet volume and 4 out of the top 10 accounts are affiliated with the Trump campaign in both months.

By volume, tweets from Trump's official account accounted for 62\% and 54\% of pro-Trump retweets compared to 68\% and 52\% for Clinton in September and October respectively.  In fact, Trump was more likely to be retweeted than Clinton.  In September, the average retweet count for Trump was 8,434 compared to 5,140 for Clinton.  Similarly in October, Trump was retweeted 13,471 time on average compared to 8,080 for Clinton.  Figures \ref{fig:VolumeTweetsCounts} and \ref{fig:RetweetsVolumeOct} show the volume of retweets per author for September and October respectively.  Tables \ref{table:TopViralTweets} and \ref{table:TopViralTweetsOct} list the most retweeted tweets over the months of September and October respectively in support of both candidates.  Again, tweets by the official accounts of the candidates dominate the top spots.  Also, aside from Trump's tweet stating ``Mexico will pay for the wall!'', none of the top tweets discuss policy, but are rather attacks against the other candidate or their supporters or self promotion.


\section{Limitations}

To better understand the results presented here, there are a few limitations that need to be considered:
\begin{enumerate}
\item The top 50 viral tweets do not have to be representative of the whole collection. Nonetheless, they still represents over 50\% of the tweets volume on the US elections during the period of the study.
\item Results are based on tweets collected from TweetElect. Although it is highly robust, the site uses automatic filtering methods that are not perfect. Therefore, there might be other relevant viral tweets that were not captured by the filtering method.
\item Measuring support for a candidate using viral tweets does not have to represent actual support on the ground for many reasons.  Some of these reasons include the fact that demographics of Twitter users may not match the general public, more popular accounts have a better chance of having their tweets go viral, or either campaign may engage in astroturfing, in which dedicated groups may methodically tweet or retweet pro-candidate messages.
\end{enumerate}



\section{Conclusion and Future Work}
\label{sec:conclusion}
In this paper, we presented the top retweeted tweets (viral tweets) from September and October 2016 supporting the two main candidates for the 2016 US presidential elections.  We provided quantitative and qualitative analysis of these top tweets.  Our results show that pro-Trump tweets were retweeted more often than pro-Clinton tweets, and he had more support on most days, especially in October, few days before the election day.

Further analysis is required to determine the leanings of the authors of negative/positive tweets and the topics of interest that they discussed. In addition, it is very interesting to study if these trends for both candidates are natural, or if campaign are engaged in astroturfing, in which dedicated groups of people methodically tweet or retweet pro-candidate messages.
\bibliographystyle{aaai}
\bibliography{ref}





\begin{table*}[t!]
\small
\center
\caption{Most retweeted accounts supporting the two candidates for September}
\label{table:TopViralTweeps}
\begin{tabular}{l|r|r|l}
\hline
\multicolumn{4}{c}{pro-Clinton}\\\hline
Account & Number of Tweets & Retweet volume & description \\\hline
Hillary Clinton & 386 & 1,984,078 & Official account \\
Ozzyoncé & 1 & 152,756 & Twitter user \\
Jerry Springer & 1 & 150,872 & Tabloid TV show presenter \\
Senator Tim Kaine & 8 & 31,769 & Running mate official account\\ 
Daniel Dale & 5 & 29,885 & Canadian journalist\\
David Fahrenthold & 4 & 29,557 & Washington Post reporter\\
Little Miss Flint & 1 & 28,863 & Civic activist\\
Bernie Sanders & 5 & 28,513 & Democratic presidential candidate\\
Joe Biden & 4 & 25,300 & Vice president\\
I'm 5'13 & 1 & 22,306 & Twitter user\\
\hline
\multicolumn{4}{c}{pro-Trump}\\\hline
Account & Number of Tweets & Retweet volume & description \\
\hline
Donald J. Trump & 251 & 2,117,249 & Official account\\
Kellyanne Conway & 48 & 106,796 & Campaign strategist\\
WikiLeaks & 22 & 102,242 & Site that released Clinton emails\\
Donald Trump Jr. & 22 & 63,611 & Trump's son\\
Paul Joseph Watson & 29 & 60,460 & Right-leaning journalist\\
Daniel Scavino Jr. & 15 & 46,300 & Campaign strategist\\
Mike Cernovich & 25 & 37,317 & Lawyer and author \\
DEPLORABLE TRUMPCAT & 14 & 29,381 & Twitter user \\
Fox News & 14 & 28,706 & Right-leaning media \\
Lou Dobbs & 9 & 22,752 & Fox News anchor\\
\hline
\end{tabular}
\end{table*}

\begin{table*}[t!]
\small
\center
\caption{Most retweeted accounts supporting the two candidates for October}
\label{table:TopViralTweepsOct}
\begin{tabular}{l|r|r|l}
\hline
\multicolumn{4}{c}{pro-Clinton}\\\hline
Account & Number of Tweets & Retweet volume & description \\\hline
Hillary Clinton	&	246	&	1987701	&	Official account	\\
Stephen King	&	2	&	135744	&	Famous author	\\
Bailey Disler	&	1	&	124322	&	Twitter user	\\
Kat Combs	&	1	&	105118	&	Twitter user	\\
Es un racista	&	1	&	102063	&	Twitter user	\\
Erin Ruberry	&	1	&	72167	&	Twitter user	\\
Senator Tim Kaine	&	12	&	68898	&	Running mate official account	\\
Richard Hine	&	1	&	66817	&	Marketing executive	\\
billy eichner	&	2	&	66761	&	Comedian and actor	\\
Jeb Bush	&	1	&	64545	&	Republican politician	\\
\hline
\multicolumn{4}{c}{pro-Trump}\\\hline
Account & Number of Tweets & Retweet volume & description \\
\hline
Donald J. Trump	&	366	&	4930413	&	Official account	\\
WikiLeaks	&	66	&	758814	&	Site that released Clinton emails	\\
Kellyanne Conway	&	60	&	271857	&	Campaign strategist	\\
Official Team Trump	&	30	&	233680	&	Trump campaing	\\
Mike Pence	&	36	&	216982	&	Running mate official account	\\
Dan Scavino Jr.	&	45	&	200315	&	Campaign strategist	\\
Paul Joseph Watson	&	40	&	189277	&	Right leaning journalist	\\
Jared Wyand ðŸ‡ºðŸ‡¸	&	20	&	110943	&	Self proclaimed nationalist	\\
Rob Fee	&	1	&	99401	&	Comedian	\\
Donald Trump Jr.	&	18	&	77036	&	Trump's son	\\

\hline
\end{tabular}
\end{table*}

\begin{table*}[t!]
\caption{Top retweeted tweets supporting each candidate for September 2016.}
\label{table:TopViralTweets}
\small
\begin{tabular}{l|l|l|r}
\hline
\multicolumn{4}{c}{Most retweeted pro-Clinton Tweets} \\
\hline
	Author & Date & Tweet & Count \\
    \hline
Ozzyoncé & 9/15	& \begin{minipage}[t]{1.4\columnwidth} Donald Trump said pregnancy is very inconvenient for businesses like his mother's pregnancy hasn't been inconvenient for the whole world. \end{minipage}	& 152,756\\
Jerry Springer & 9/27	& \begin{minipage}[t]{1.4\columnwidth} Hillary Clinton belongs in the White House. Donald Trump belongs on my show.\end{minipage}	& 150,872\\
Hillary Clinton	& 9/27 & \begin{minipage}[t]{1.4\columnwidth} RT this if you re proud to be standing with Hillary tonight. \#debatenight https://t.co/91tBmKxVMs \end{minipage}	 & 72,443\\
Hillary Clinton	& 9/27 & \begin{minipage}[t]{1.4\columnwidth} I never said that. -- Donald Trump who said that. \#debatenight https://t.co/6T8qV2HCbL \end{minipage}	& 72,111\\
Hillary Clinton & 9/10 &	\begin{minipage}[t]{1.4\columnwidth} Except for African Americans Muslims Latinos immigrants women veterans -- and any so-called losers or dummies. https://t.co/rbBg2rXZdm  \end{minipage} &	50,913\\
\hline
\hline
\multicolumn{4}{c}{Most retweeted pro-Trump Tweets} \\
\hline
	Author & Date & Tweet & Count \\
    \hline
Donald J. Trump & 9/24	& \begin{minipage}[t]{1.4\columnwidth} If dopey Mark Cuban of failed Benefactor fame wants to sit in the front row perhaps I will put Gennifer Flowers right alongside of him \end{minipage}!	& 30,950\\
Donald J. Trump	& 9/20 & \begin{minipage}[t]{1.4\columnwidth} Hillary Clinton is taking the day off again she needs the rest. Sleep well Hillary - see you at the debate! \end{minipage} & 30,900\\
Donald J. Trump	& 9/01 & \begin{minipage}[t]{1.4\columnwidth} Mexico will pay for the wall!  \end{minipage} &	30,464\\
Donald J. Trump & 9/27 & \begin{minipage}[t]{1.4\columnwidth} Nothing on emails. Nothing on the corrupt Clinton Foundation. And nothing on \#Benghazi. \#Debates2016 \#debatenight \end{minipage} &	27,177\\
Donald J. Trump & 9/10 & \begin{minipage}[t]{1.4\columnwidth} While Hillary said horrible things about my supporters and while many of her supporters will never vote for me I still respect them all!	\end{minipage} & 25,882
\\
\hline
\end{tabular}
\end{table*}

\begin{table*}[t!]
\caption{Top retweeted tweets supporting each candidate for October 2016.}
\label{table:TopViralTweetsOct}
\small
\begin{tabular}{l|l|l|r}
\hline
\multicolumn{4}{c}{Most retweeted pro-Clinton Tweets} \\
\hline
	Author & Date & Tweet & Count \\
    \hline
Bailey Disler & 10/26	& \begin{minipage}[t]{1.4\columnwidth} Good morning everyone especially the person who destroyed Donald Trump's walk of fame star https://t.co/IcBthxMPd9
 \end{minipage}	& 124,322\\
Stephen King & 10/21	& \begin{minipage}[t]{1.4\columnwidth} My newest horror story: Once upon a time there was a man named Donald Trump and he ran for president. Some people wanted him to win.\end{minipage}	& 121,635 \\
Kat Combs & 10/10 & \begin{minipage}[t]{1.4\columnwidth} Trump writing a term paper: Sources Cited: 1. You Know It 2. I know It 3. Everybody Knows It \end{minipage}	 & 105,118\\
Es un racista & 10/08 & \begin{minipage}[t]{1.4\columnwidth} Anna for you to sit here call Trump a racist is outrageous Anna: OH?! Well lemme do it again in 2 languages! https://t.co/nq4DO7bN7J
 \end{minipage}	& 102,063\\
Rob Fee & 10/08 & \begin{minipage}[t]{1.4\columnwidth} How are so many people JUST NOW offended by Trump? It s like getting to the 7th Harry Potter book realizing Voldemort might be a bad guy. \end{minipage} &	99,401\\
\hline
\hline
\multicolumn{4}{c}{Most retweeted pro-Trump Tweets} \\
\hline
	Author & Date & Tweet & Count \\
    \hline
Donald J. Trump & 10/08 & \begin{minipage}[t]{1.4\columnwidth} Here is my statement. https://t.co/WAZiGoQqMQ \end{minipage} &	52,887\\
WikiLeaks & 10/14 & \begin{minipage}[t]{1.4\columnwidth}	Democrats prepared fake Trump grope under the meeting table Craigslist employment advertisement in May 2016… https://t.co/JM9JMeLYet \end{minipage} & 45,348\\
WikiLeaks & 10/03 & \begin{minipage}[t]{1.4\columnwidth} Hillary Clinton on Assange Can't we just drone this guy -- report https://t.co/S7tPrl2QCZ https://t.co/qy2EQBa48y  \end{minipage} & 45,233\\
Mike Pence & 10/10 & \begin{minipage}[t]{1.4\columnwidth} Congrats to my running mate @realDonaldTrump on a big debate win! Proud to stand with you as we \#MAGA. \end{minipage} & 42,178\\
Donald J. Trump & 10/08 & \begin{minipage}[t]{1.4\columnwidth} The media and establishment want me out of the race so badly - I WILL NEVER DROP OUT OF THE RACE WILL NEVER LET MY SUPPORTERS DOWN! \#MAGA \end{minipage} &	41,386
\\
\hline
\end{tabular}
\end{table*}

\end{document}